\documentclass{iau} 
\usepackage{graphicx}

\title[Symp. 346.~~Magnetic fields in SgHMXBs] 
{Studying the presence of magnetic fields in a sample of high-mass X-ray binaries}

\author[S.~Hubrig et al.]   
{Swetlana~Hubrig$^1$,
Alexander~F.~Kholtygin$^2$,
Lara~Sidoli$^3$,
Markus~Sch{\"o}ller$^4$,
\and Silva~P.~J{\"a}rvinen$^1$}

\affiliation{$^1$Leibniz-Institut f\"ur Astrophysik Potsdam (AIP), 
An der Sternwarte~16, 14482~Potsdam, Germany, email: {\tt shubrig@aip.de} \\[\affilskip]
$^2$Saint-Petersburg State University, Universitetskij pr.~28, 198504~Saint-Petersburg, Russia\\
$^3$INAF, Istituto di Astrofisica Spaziale e Fisica Cosmica, Via E.~Bassini~15, 20133 Milano, Italy\\
$^4$European Southern Observatory, Karl-Schwarzschild-Str.~2, 85748~Garching, Germany
}

\pubyear{2018}
\volume{346}  
\jname{High-mass X-ray binaries:
illuminating the passage from massive binaries
to merging compact objects}
\editors{A.C. Editor, B.D. Editor \& C.E. Editor, eds.}
\begin{document}

\maketitle

\begin{abstract}
Previous circular polarization observations obtained with the ESO FOcal Reducer low dispersion 
spectrograpgh 
at the VLT in 2007--2008 revealed the presence of a weak longitudinal 
magnetic field on the surface of the optical component of the X-ray binary 
Cyg X-1, which contains a black hole and an O9.7Iab supergiant on a 5.6\,d orbit.
In this contribution we report on recently acquired FORS\,2 spectropolarimetric 
observations of Cyg X-1 along with measurements of a few additional high-mass 
X-ray binaries. 
\end{abstract}

\firstsection 

\section{Introduction}

High-mass X-ray binaries are fundamental for studying stellar evolution,
nucleosynthesis, structure and evolution of galaxies, and accretion processes. 
The classical high-mass X-ray binaries (HMXBs) with supergiant companions (SgHMXBs) are known since 
the birth of X-ray astronomy and are persistent X-ray emitters, with a limited range of intensity 
variability (around a factor of 10). These targets are among 
the brightest X-ray sources in the sky. For them, the observed spectral and time variability is 
best explained by assuming that accretion onto the compact object is taking place from a highly 
structured stellar wind, where cool dense clumps are embedded in a rarefied photoionized gas.

While the first spectropolarimetric observations using the FOcal Reducer low dispersion 
Spectrograph (FORS\,1/2; Appenzeller et al.\ 1998) mounted on the 8\,m Antu telescope of 
the Very Large Telescope indicated the presence of a rather strong longitudinal magnetic field of a kG order in the  
supergiant fast X-ray transient (SFXT)  IGR\,J11215-5952 (Hubrig et al.\ 2018), no 
systematical search for magnetic fields
was carried out in SgHMXBs.
SFXTs are a subclass of HMXBs associated with early-type
supergiant companions, and characterized by sporadic, short and bright X–ray flares
reaching peak luminosities of 10$^{36}$--10$^{37}$\,erg\,s$^{-1}$ and typical energies released in 
bright flares of about 10$^{38}$--10$^{40}$\,erg (see the review by Sidoli in 2017 for more details). 
Different accretion regimes - transient in the 
settling accretion mode versus persistent in the free-fall Bondi mode -
were suggested in the last years for SFXTs and SgHMXBs, respectively.

The magnetic field of the eclipsing binary Vela\,X-1 with a B0.5Ia component 
and an orbital period of 8.96\,d 
was previosly studied by Hubrig et al.\ (2013). However, no magnetic field
detection at a significance level of 3$\sigma$ has been achieved in this system.
The spectral behaviour of  Vela\,X-1 is known to be very complex due to the presence of bumps and
wiggles in the line profiles, and an impact of tidal effects producing  orbital  phase-dependent  
variations in the line profiles
leading to asymmetries such as extended blue or red wings (Koenigsberger et al.\ 2012).
 Furthermore, Kreykenbohm et al.\ (2008) detected flaring activity and temporary quasi-periodic 
oscillations in INTEGRAL X-ray observations.
Magnetic field measurements of the X-ray binary Cyg\,X-1, with the historically first
black-hole candidate, using FORS\,2 low-resolution
spectropolarimetric observations were reported by Karitskaya et al.\ (2010).
The authors detected a relatively weak mean longitudinal magnetic fields of the order of 100\,G with a few
measurements at a significance level in the range between 3.5 and 6.2$\sigma$.

In this contribution we discuss the most recent FORS\,2 spectropolarimetric observations 
of the Cyg\,X-1 system and two other SgHMXB systems, BP\,Cru, and LS\,5039.
For completeness we also present the older results for 
Vela\,X-1, as this system can be considered as the prototype of persistent HMXBs.

\section{Magnetic field measurements}

\begin{figure}
 \centering 
        \includegraphics[width=0.75\textwidth]{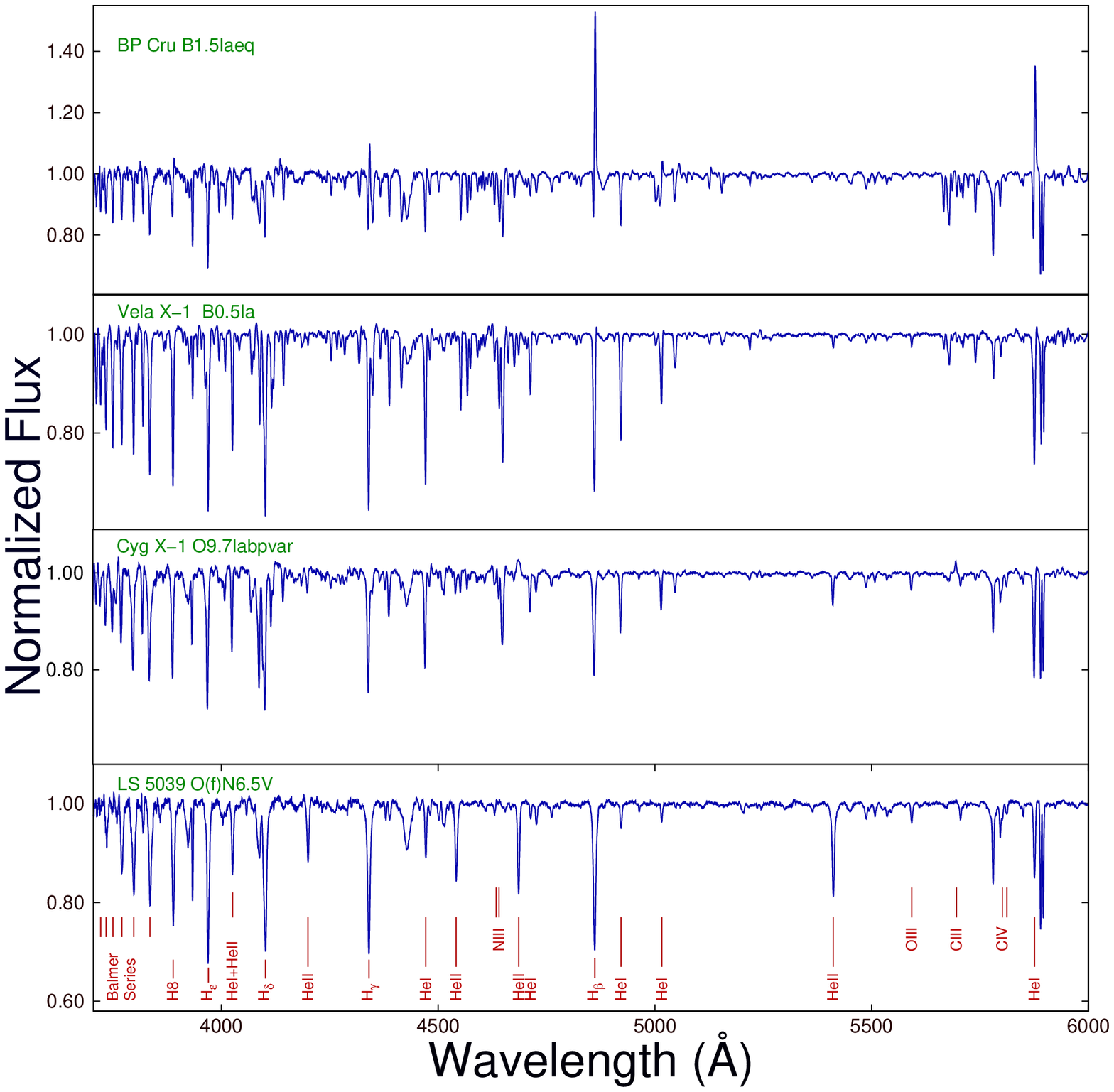}
        \caption{
          Normalised FORS\,2 spectra of LS\,5039, Cyg\,X-1, Vela\,X-1, and BP\,Cru. Well known spectral
lines are indicated.
         }
   \label{fig:all}
\end{figure}

The FORS\,2 multi-mode instrument is equipped with polarisation analysing optics
comprising super-achromatic half-wave and quarter-wave phase retarder plates,
and a Wollaston prism with a beam divergence of 22$^{\prime\prime}$ in standard
resolution mode. 
We used the GRISM 600B and the narrowest available slit width
of 0.4$^{\prime\prime}$ to obtain a spectral resolving power of $R\approx2000$.
The observed spectral range from 3250 to 6215\,\AA{} includes all Balmer lines,
apart from H$\alpha$, and numerous helium lines.
For the observations, we used a non-standard readout mode with low 
gain (200kHz,1$\times$1,low), which provides a broader dynamic range, hence 
allowed us to reach a higher signal-to-noise ratio in the individual spectra.
The  spectral  appearance  of all targets in the FORS\,2 spectra
is presented in Fig.~\ref{fig:all}.

The longitudinal magnetic field was measured in two ways: using the entire spectrum
including all available lines
or using exclusively the hydrogen lines.
Furthermore, we carried out Monte Carlo bootstrapping tests. 
These are most often applied with the purpose of deriving robust estimates of standard errors. 
The measurement uncertainties obtained before and after the Monte Carlo bootstrapping tests were found to be 
in close agreement, indicating the absence of reduction flaws. 
The results of our magnetic field measurements, those for the entire spectrum
or only for the hydrogen lines, are presented in Table~\ref{tab:priMFFORS}.

\begin{table}
\centering
\caption{Longitudinal magnetic field values obtained using FORS\,2 observations of four high-mass 
X-ray binaries. In the first two columns we show the name of the binary and the modified 
Julian date of mid-exposure, followed by the mean longitudinal magnetic field using the
  Monte Carlo bootstrapping test, for all lines and for the hydrogen lines. In the
  last column, we present the significance of the measurements of
  $\left<B_{\rm z}\right>_{\rm all}$ using the set of all lines. All quoted
  errors are 1$\sigma$ uncertainties. 
}
\label{tab:priMFFORS}
\begin{tabular}{ccr@{$\pm$}lr@{$\pm$}lc}
\hline
\hline\\
\multicolumn{1}{c}{Name} &
\multicolumn{1}{c}{MJD} &
\multicolumn{2}{c}{$\left<B_{\rm z}\right>_{\rm all}$} &
\multicolumn{2}{c}{$\left<B_{\rm z}\right>_{\rm hyd}$} &
\multicolumn{1}{c}{Significance} \\
&
&
\multicolumn{2}{c}{[G]} &
\multicolumn{2}{c}{[G]} &
\multicolumn{1}{c}{$\sigma$}\\
\hline
BP\,Cru &57528.0781     & 344 & 149 & 441 & 173 & 2.3 \\
BP\,Cru &57533.1965     & $-$144 & 162 & $-$319 & 193 & 0.9 \\
BP\,Cru &57591.0529     & 254 & 162 & 82 &214 & 1.6 \\
Cyg\,X-1&57585.1653     & 159    & 63 & 447    & 221 & 2.5 \\
Cyg\,X-1&57593.2230     & $-$61   & 82 &$-$ 274 & 237 & 0.8 \\
Cyg\,X-1&57644.0254     & $-$88   & 44 &$-$111  & 157 & 2.0 \\
Cyg\,X-1&57645.0189     & $-$79   & 45 &$-$52   & 114 & 1.7 \\
Vela\,X-1&55686.0962    & $-$15   & 30 &$-$11 & 45 & 0.5 \\
Vela\,X-1&55687.0584    & $-$80   & 32 &$-$114& 50 & 2.5 \\
Vela\,X-1&55688.0479    & 57      & 37 & 88   & 69 & 1.5 \\
LS\,5039&57585.1260     & 794    & 277 & 634    & 317 & 2.5 \\
LS\,5039&57591.0970     &$-$517  & 291 &$-$118 & 383 & 1.8 \\
LS\,5039&57646.1463     & 534    & 268 & 199    & 340 & 2.0 \\
LS\,5039&57611.2456     & 525    & 450 & 953    & 617 & 1.2 \\
\hline
\end{tabular}
\end{table}

\begin{figure*}
 \centering 
\includegraphics[width=0.35\textwidth]{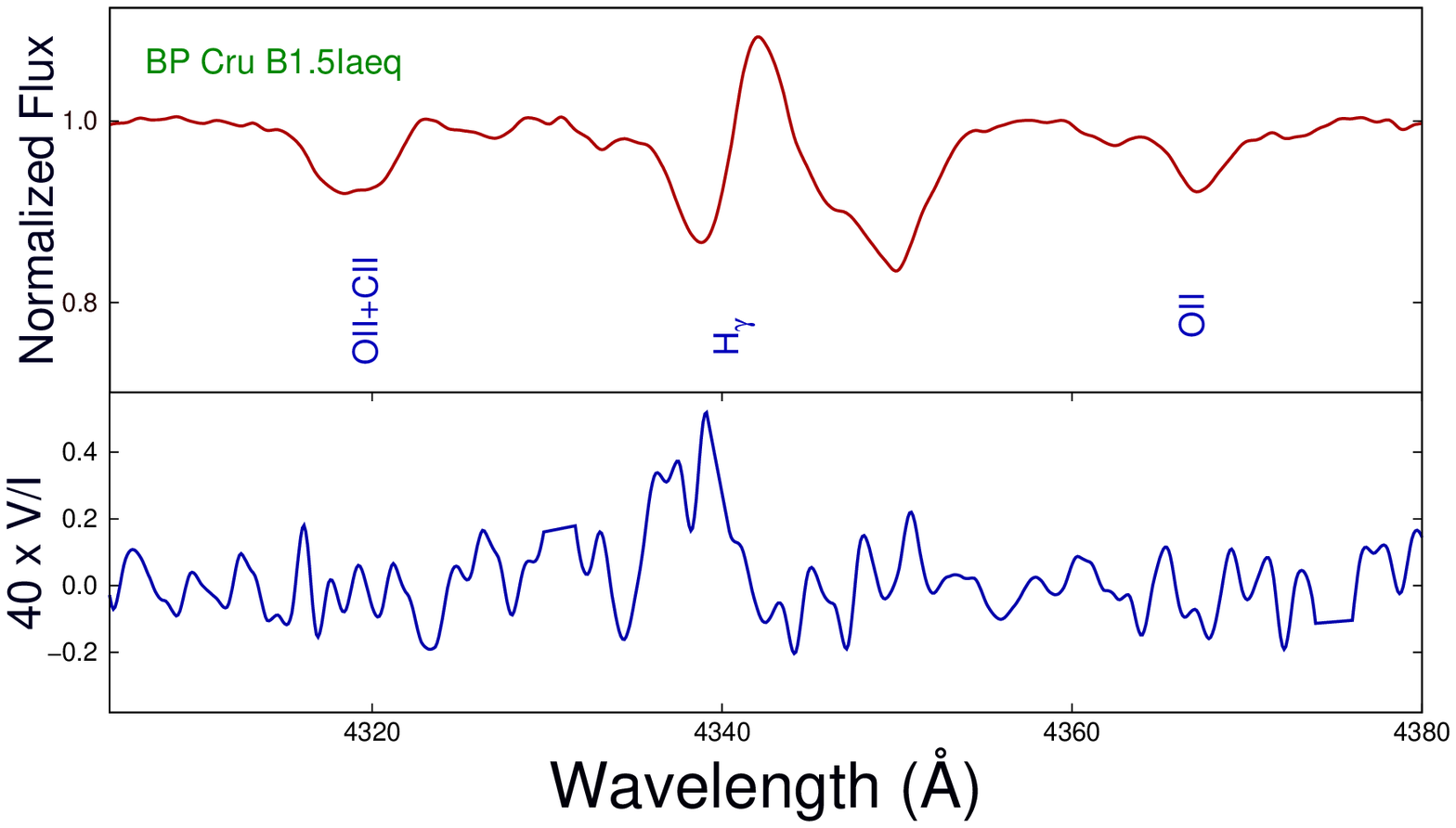}
\includegraphics[width=0.35\textwidth]{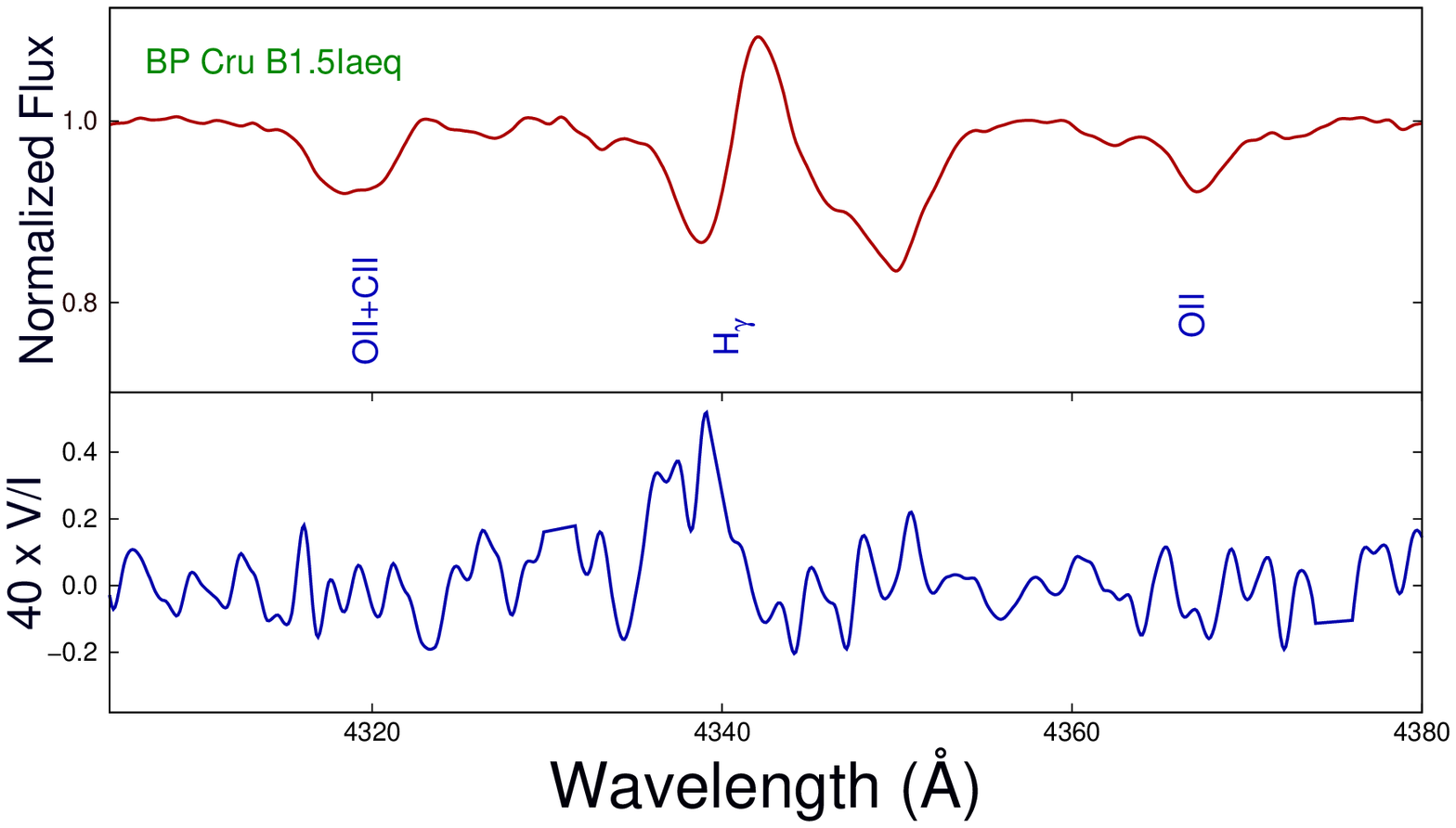}
         \caption{Examples of Stokes~$I$ and Stokes~$V$ spectra of BP\,Cru (left panel) and Cyg\,X-1 
(right panel) in the vicinity of the $H\gamma$ 
line and the He\,{\sc i}\,5876 line, respectively.
         }
   \label{fig:zf}
\end{figure*}

Our measurements do not reveal the presence of significant mean longitudinal magnetic fields
in any of the four studied binaries. We observe changes of the field polarity in 
the measurements of all targets, but the measurement uncertainties are rather large, leading
to significance levels of only 2.3--2.5$\sigma$ and less.
On the other hand, as we show in Fig.~\ref{fig:zf}, typical Zeeman features are 
detected in the FORS\,2 Stokes~$V$ spectra of BP\,Cru and Cyg\,X-1.

\section{Discussion}

\begin{figure*}
 \centering 
        \includegraphics[width=0.32\textwidth]{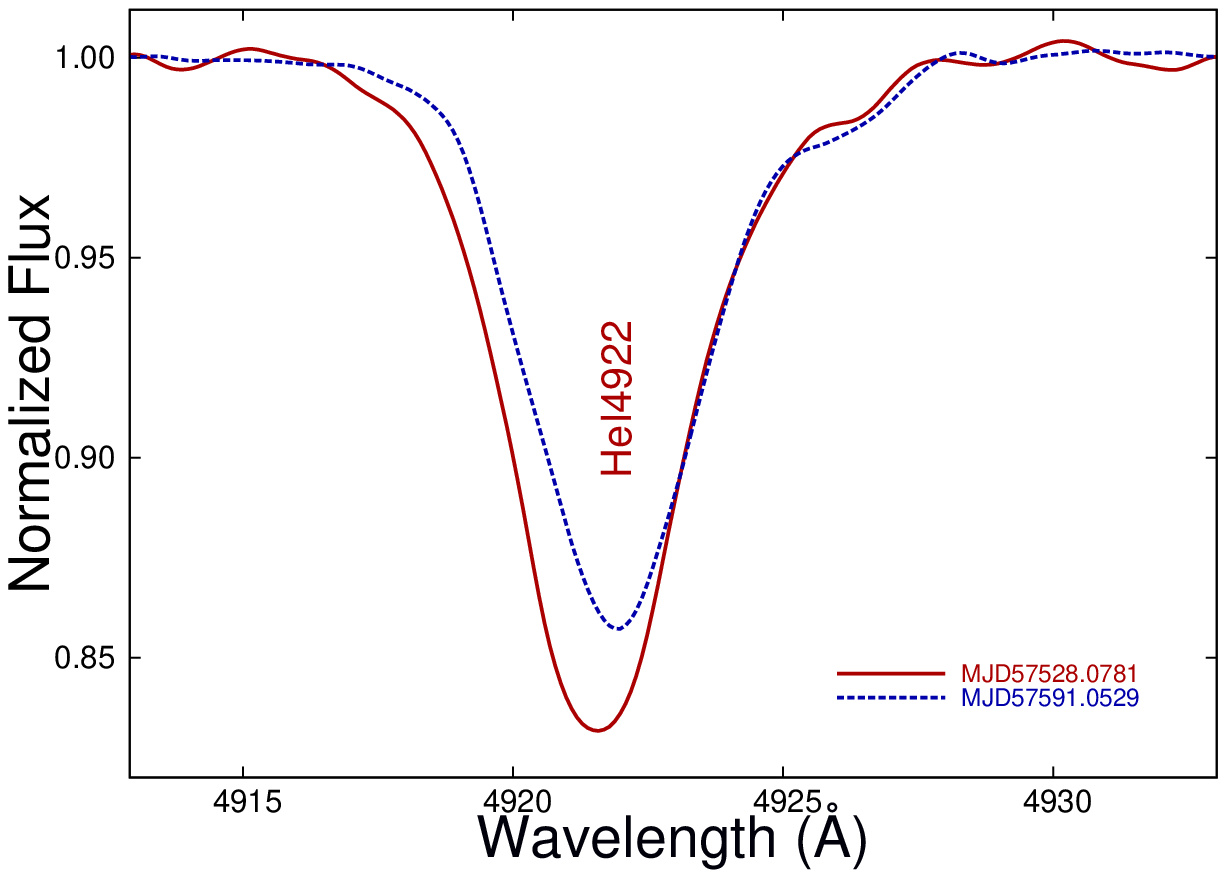}
     \includegraphics[width=0.32\textwidth]{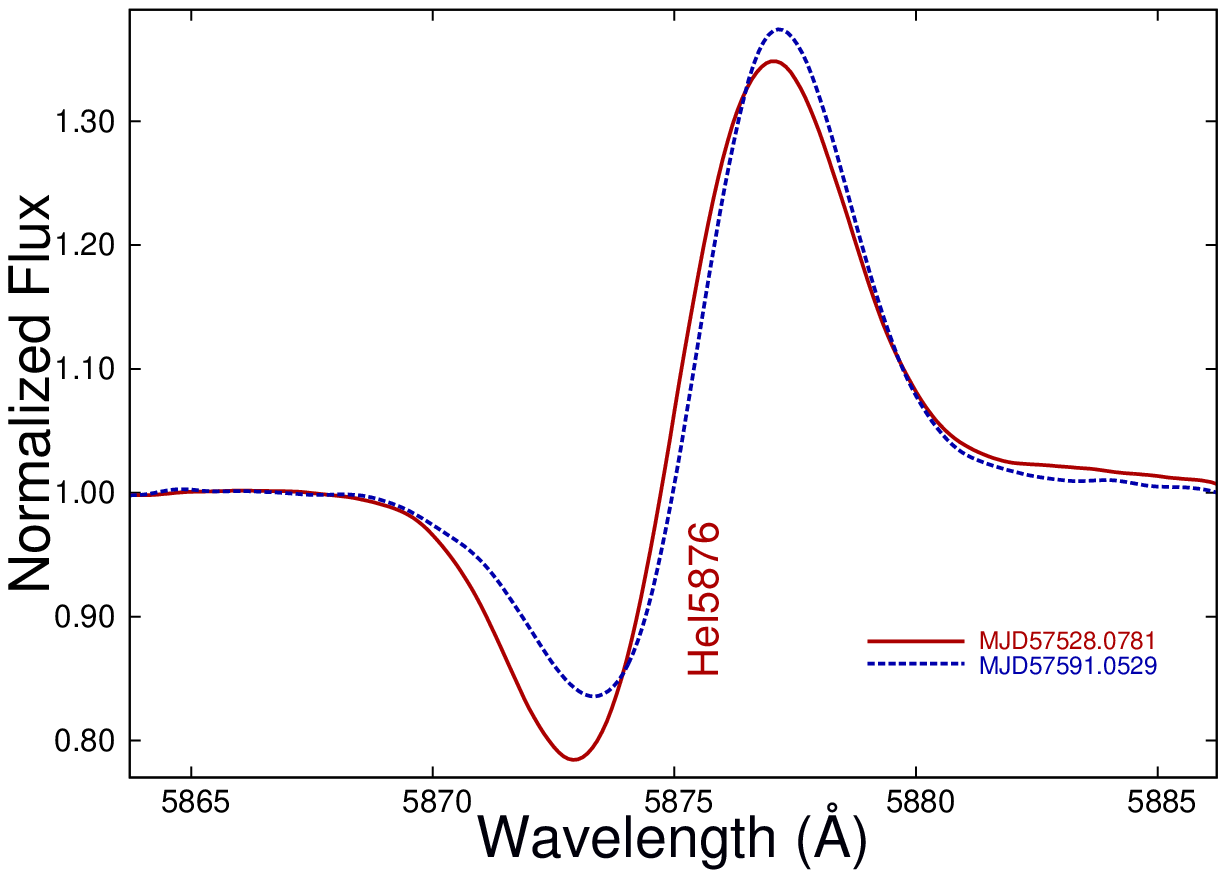}
      \includegraphics[width=0.32\textwidth]{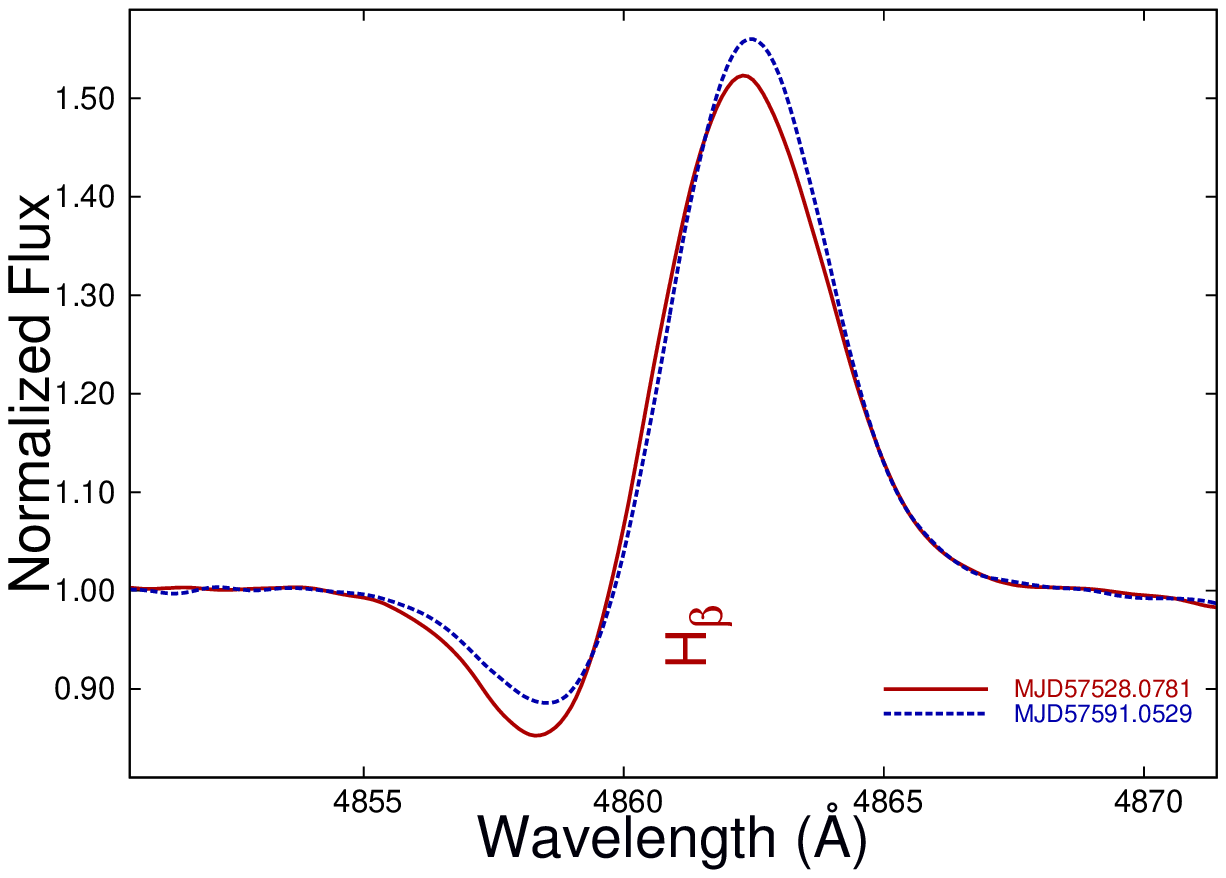}
     \caption{
 Variability of spectral lines in the FORS\,2 spectra of BP\,Cru recorded on two different nights.  
}
   \label{fig:bp}
\end{figure*}

\begin{figure*}
 \centering 
        \includegraphics[width=0.32\textwidth]{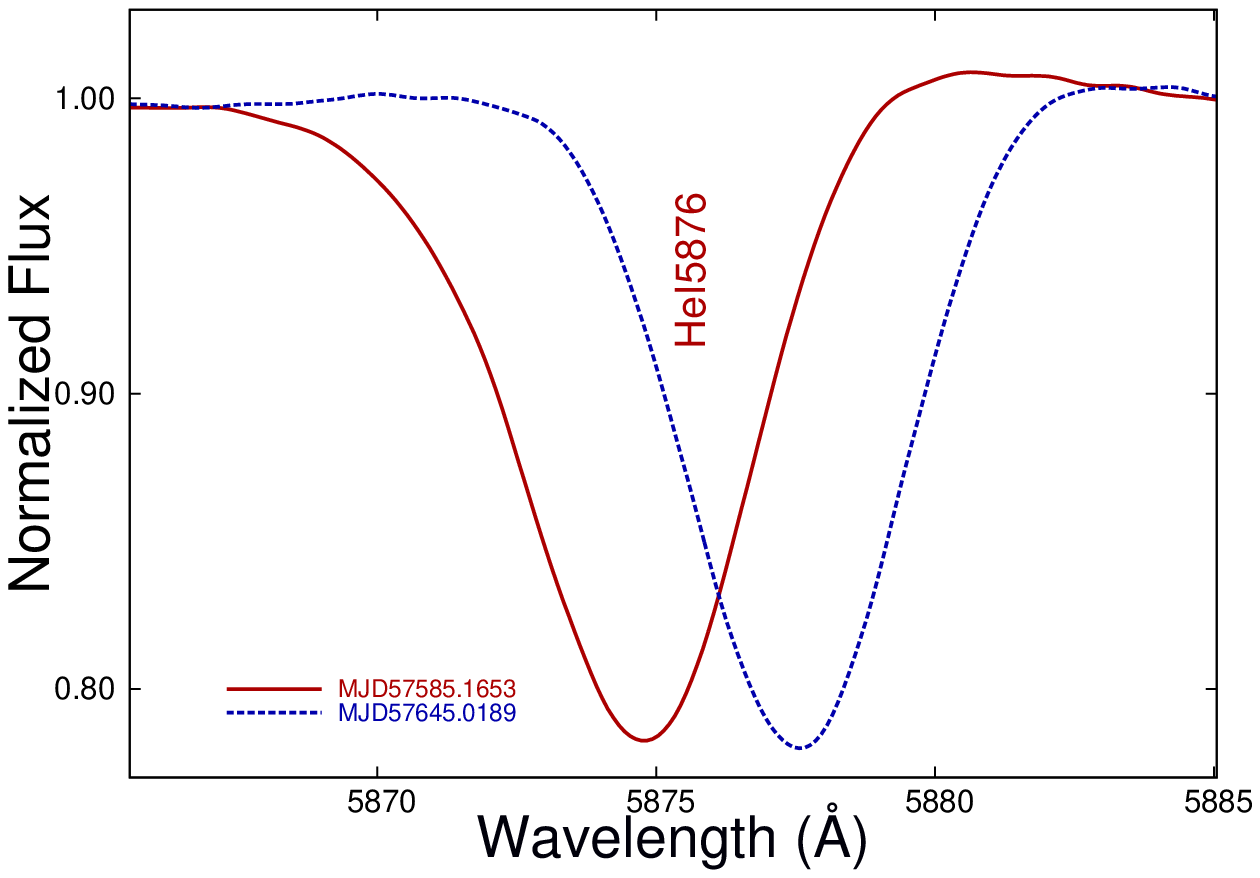}
       \includegraphics[width=0.32\textwidth]{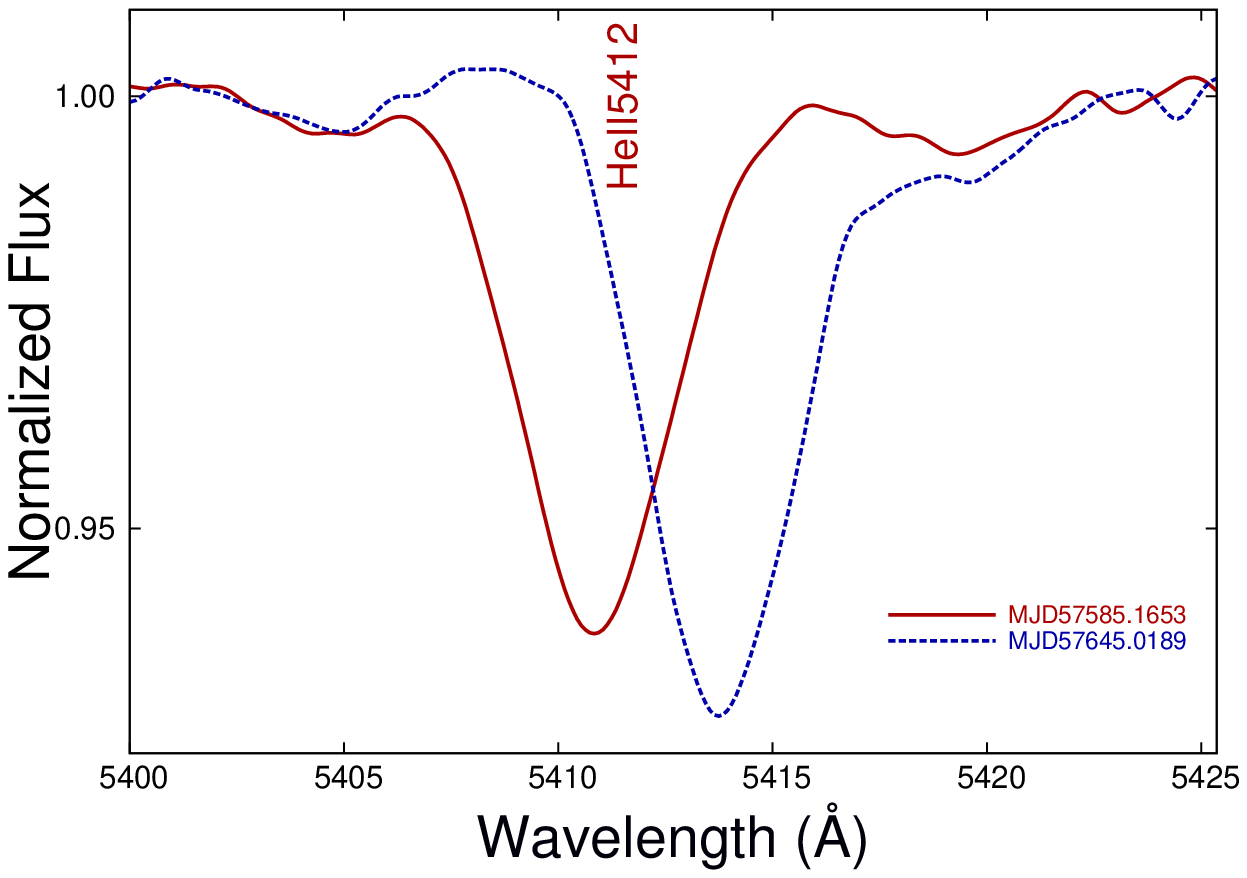}
       \includegraphics[width=0.32\textwidth]{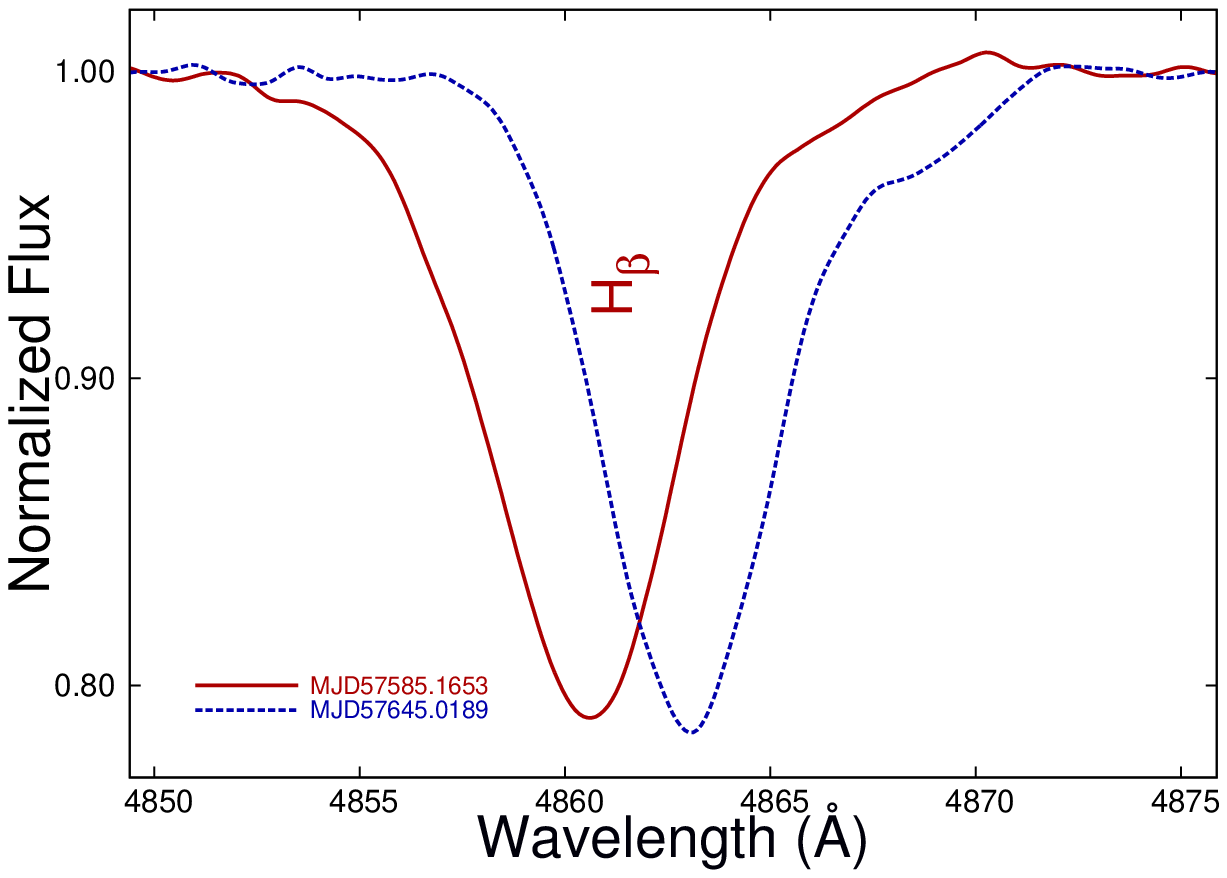}
        \caption{
Same as in Fig.~\ref{fig:bp}, but for Cyg\,X-1.
}
   \label{fig:cyg}
\end{figure*}

\begin{figure*}
 \centering 
        \includegraphics[width=0.32\textwidth]{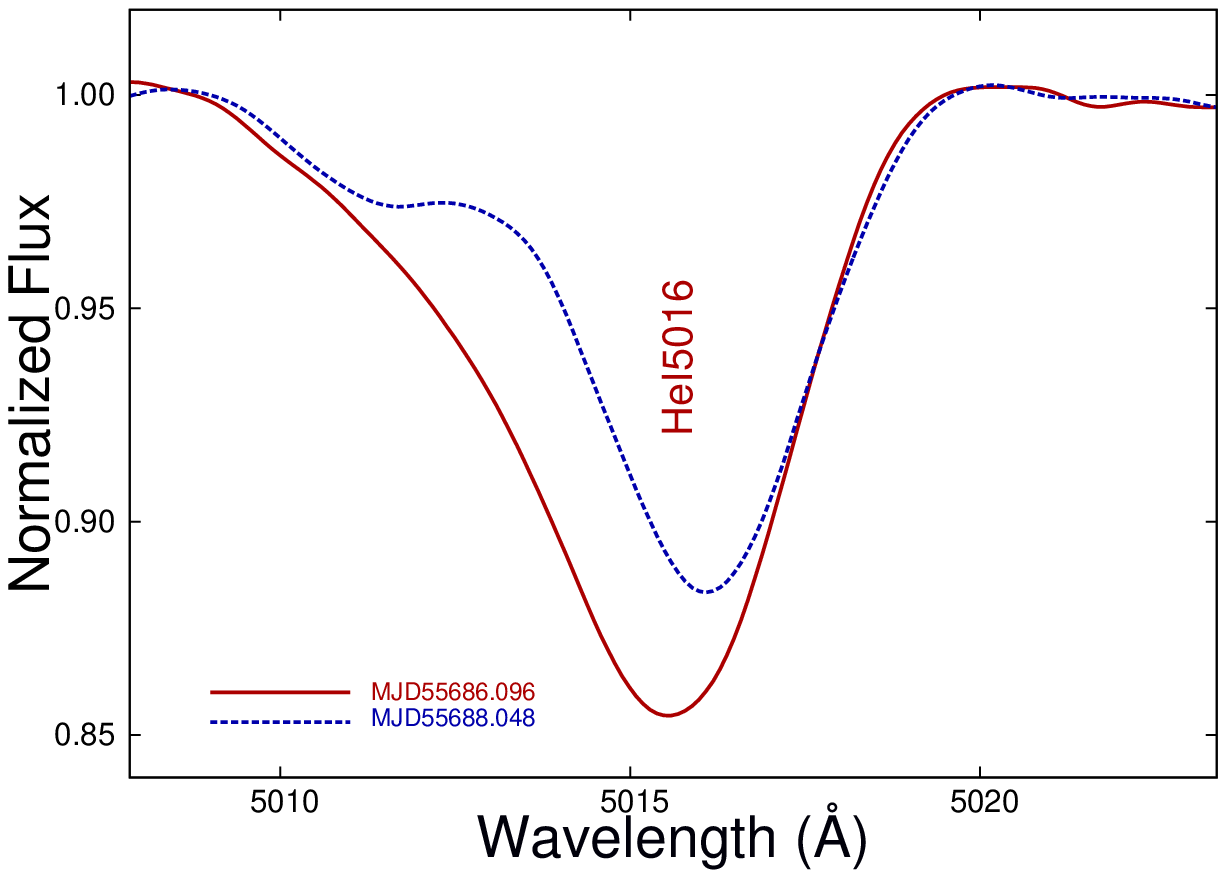}
\includegraphics[width=0.32\textwidth]{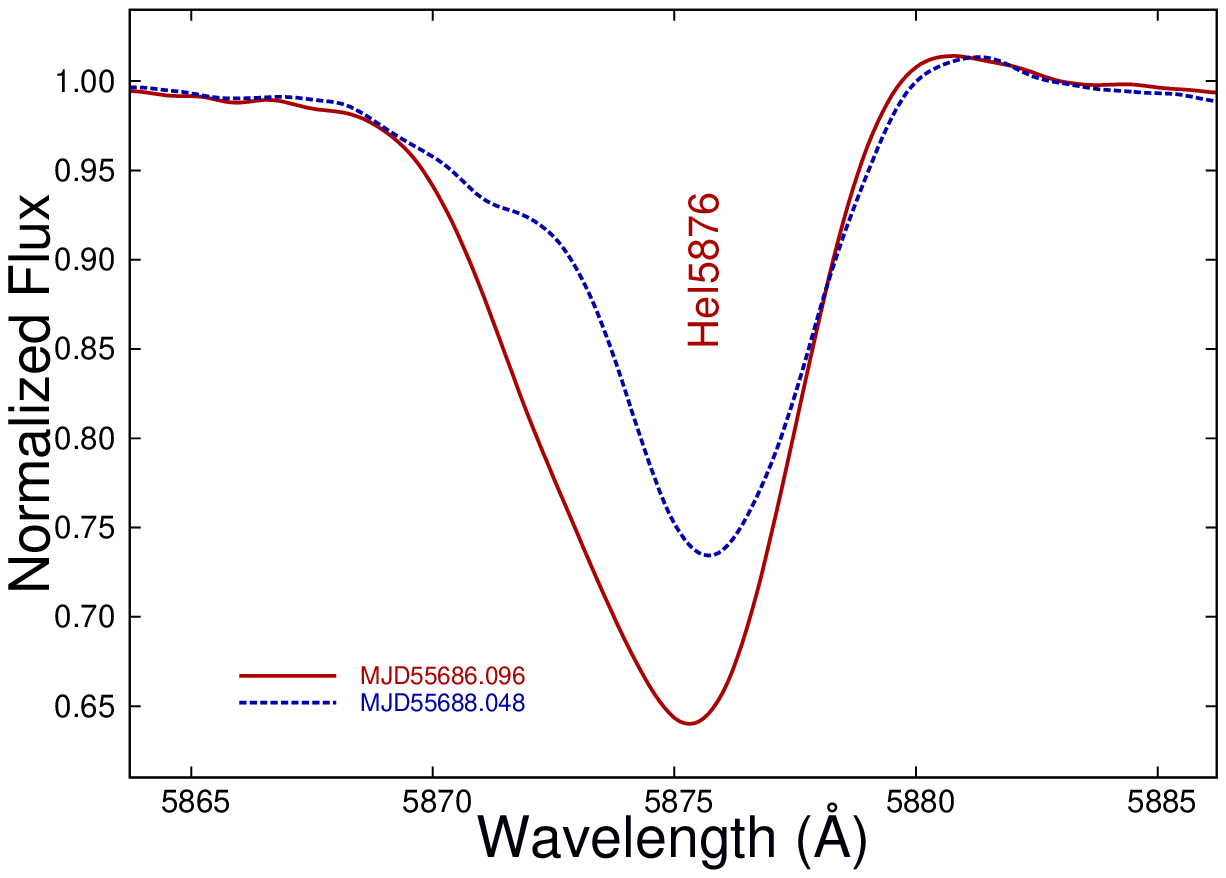}
\includegraphics[width=0.32\textwidth]{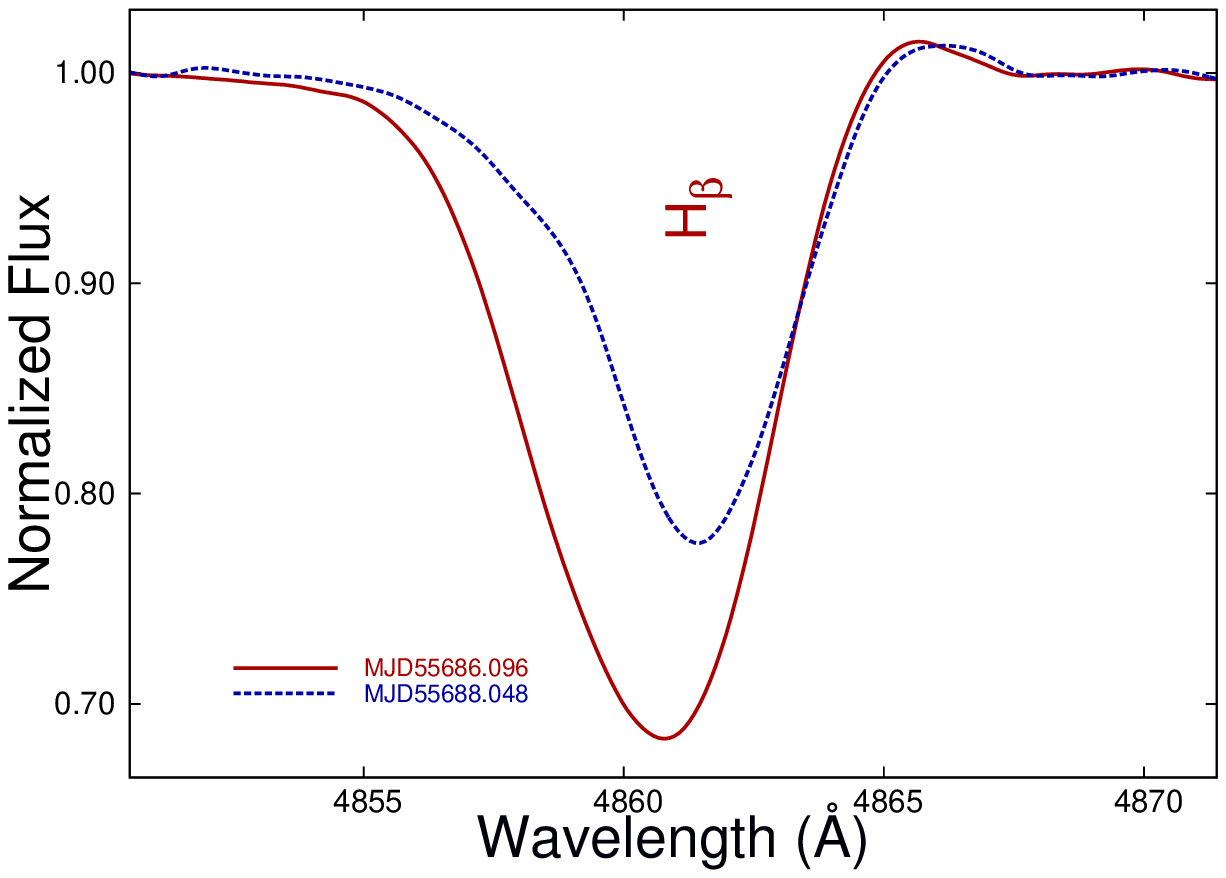}
\caption{
Same as in Fig.~\ref{fig:bp}, but for Vela\,X-1.
}
         \label{fig:vela}
\end{figure*}

\begin{figure*}
 \centering 
 \includegraphics[width=0.32\textwidth]{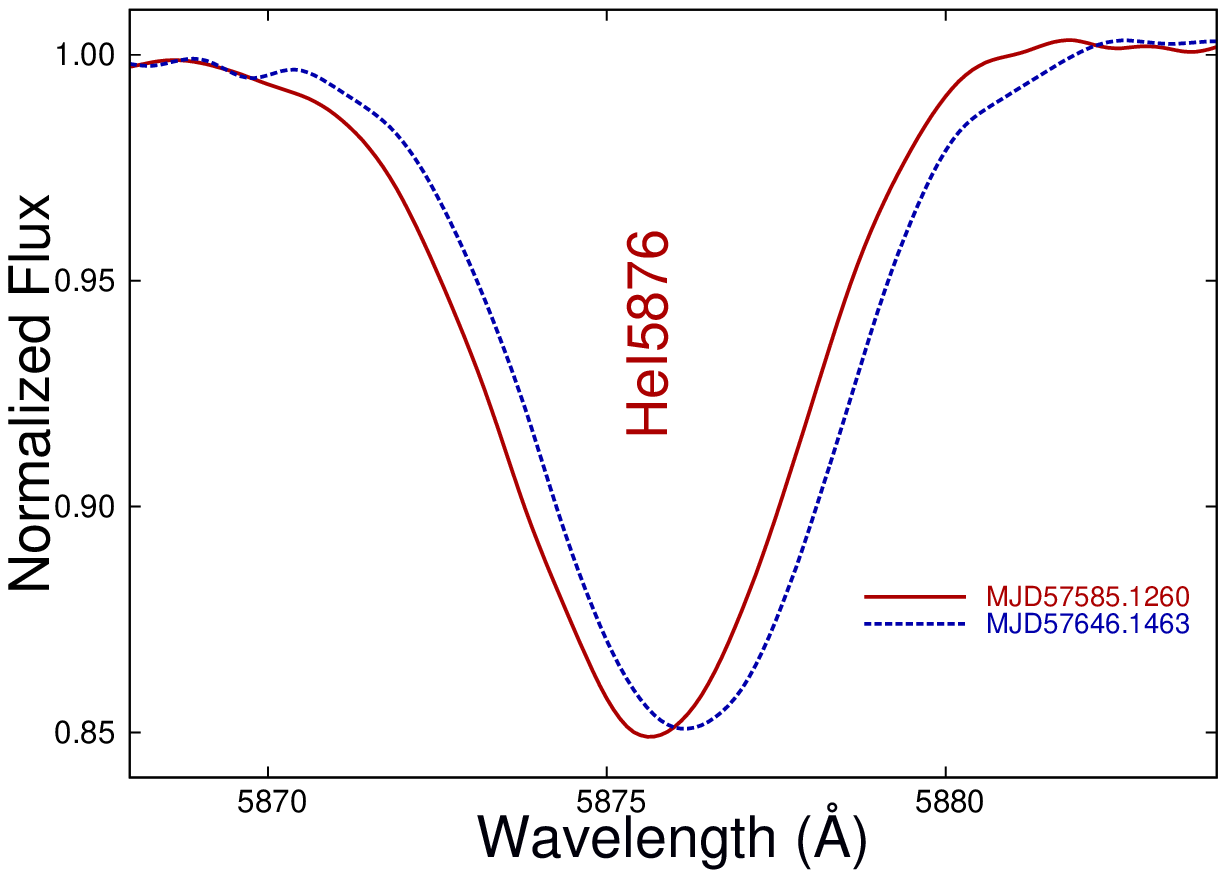}
 \includegraphics[width=0.32\textwidth]{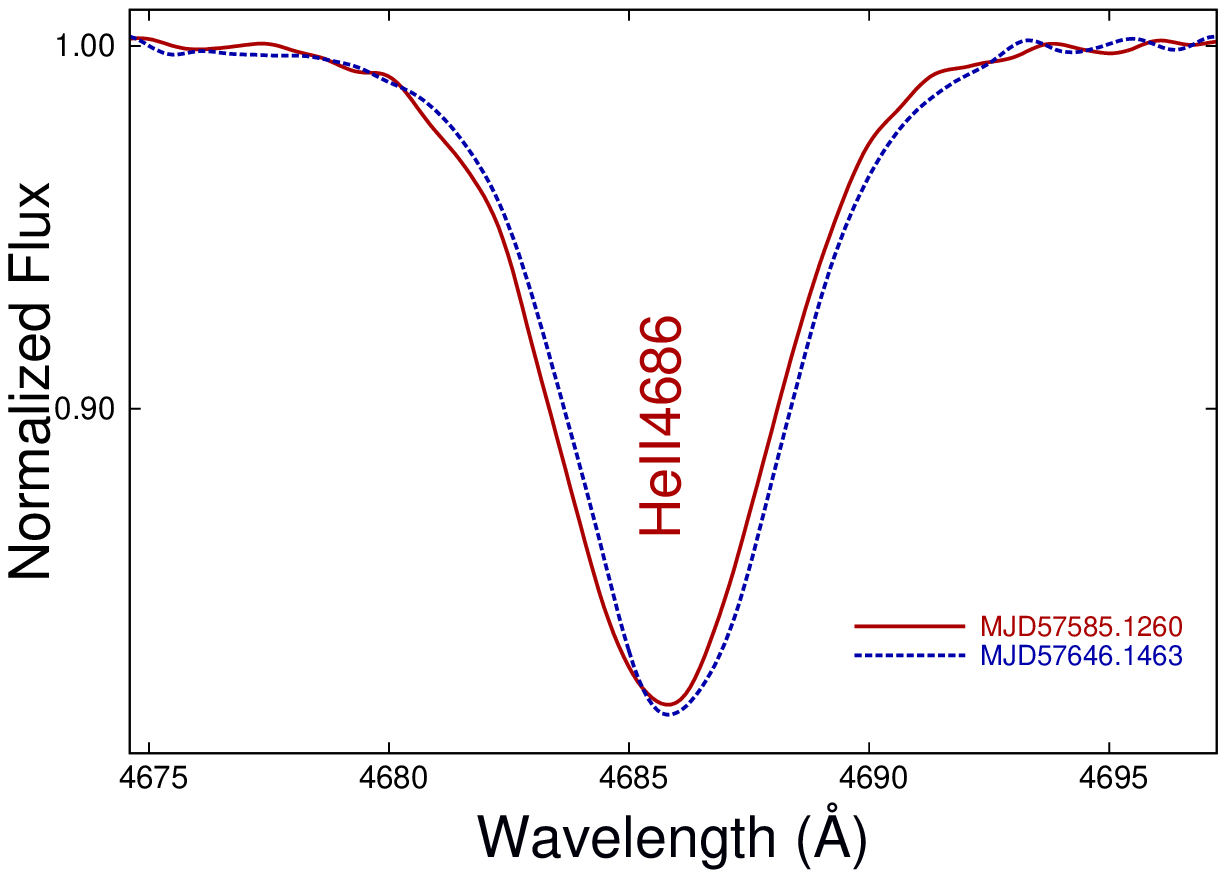}
 \includegraphics[width=0.32\textwidth]{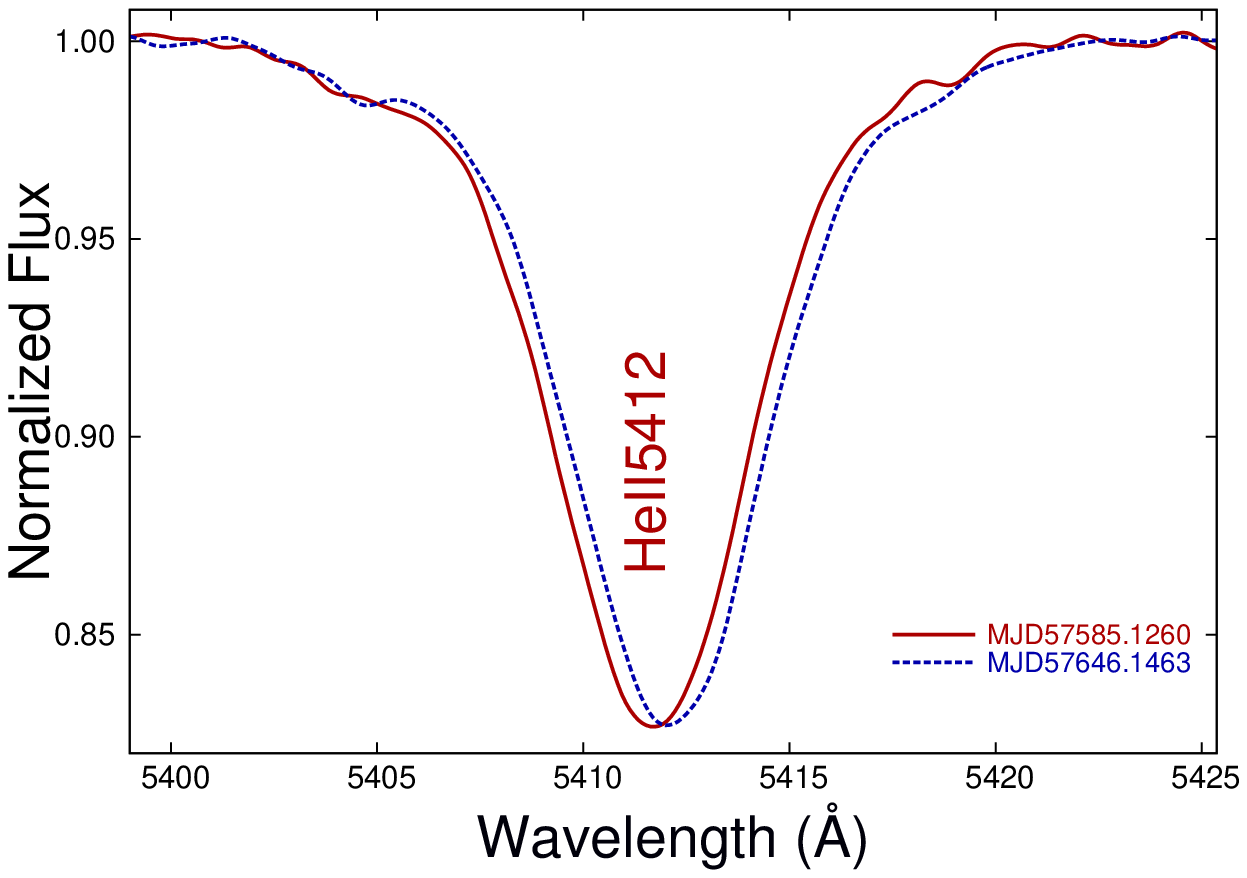}
        \caption{
Same as in Fig.~\ref{fig:bp}, but for LS\,5039.   
}
   \label{fig:LS5039}
\end{figure*}

The recent spectropolarimetric observations of four SgHMXBs with FORS\,2 showed changes in 
the field polarities,
but the measurement uncertainties are too large to allow us to conclude on the 
presence of magnetic fields. 
Apart from the search for magnetic fields, the acquired spectra of the optical components allowed us to detect
significant spectral variability:
spectral lines belonging to hydrogen and 
other elements show changes of the line intensities
and radial velocities over different observing nights. We present a few examples in 
Figs.~\ref{fig:bp}--\ref{fig:LS5039} showing
individual Stokes~$I$ helium and hydrogen line profiles. It is not clear yet whether the detected spectral 
variability is caused by the presence of magnetospheres or by pulsational variability, frequently detected
in massive OB-type stars. Future observations are urgently needed to be able to draw solid conclusions 
about the role of magnetic fields in these targets.

\end{document}